%% file: main.tex
\pdfoutput=1

\documentclass[sigconf,screen,authorversion ]{acmart}
\usepackage{fancyhdr}
\AtBeginDocument{%
    \addtolength{\footskip}{2.0\baselineskip}%
    \fancyfoot[L]{\textit{\textit{This version is a pre-print. For the definitive Version of Record, please refer to the ACM Digital Library.}}}%
}

\AtBeginDocument{%
  \providecommand\BibTeX{{%
    \normalfont B\kern-0.5em{\scshape i\kern-0.25em b}\kern-0.8em\TeX}}}

\usepackage{microtype}

\usepackage{booktabs}

\begin{document}

\setcopyright{acmlicensed}
\acmDOI{10.1145/3663531.3664752}
\acmYear{2024}
\copyrightyear{2024}
\acmSubmissionID{fsews24qsenemain-id3-p}
\acmISBN{979-8-4007-0673-8/24/07}
\acmConference[QSE-NE '24]{Proceedings of the 1st ACM International Workshop on Quantum Software Engineering: The Next Evolution}{July 16, 2024}{Porto de Galinhas, Brazil}
\acmBooktitle{Proceedings of the 1st ACM International Workshop on Quantum Software Engineering: The Next Evolution (QSE-NE '24), July 16, 2024, Porto de Galinhas, Brazil}
\received{2024-04-25}
\received[accepted]{2024-05-06}

\title{$Classi|Q\rangle$: Towards a Translation Framework to Bridge the Classical-Quantum Programming Gap}

\author{Matteo Esposito}
\orcid{0000-0002-8451-3668}
\affiliation{%
  \institution{University of Rome Tor Vergata}
  \city{Rome}
  \country{Italy}
}
\email{m.esposito@ing.uniroma2.it}

\author{Maryam Tavassoli Sabzevari}
\orcid{0009-0004-8879-0285}
\affiliation{%
  \institution{University of Oulu}
  \city{Oulu}
  \country{Finland}
}
\email{maryam.tavassolisabzevari@oulu.fi}

\author{Boshuai Ye}
\orcid{0009-0000-3480-1234}
\affiliation{%
  \institution{Aalto University}
  \city{Espoo}
  \country{Finland}
}
\email{boshuai.ye@aalto.fi}

\author{Davide Falessi}
\orcid{0000-0002-6340-0058}
\affiliation{%
  \institution{University of Rome Tor Vergata}
  \city{Rome}
  \country{Italy}
}
\email{d.falessi@gmail.com}

\author{Arif Ali Khan}
\orcid{0000-0002-8479-1481}
\affiliation{%
  \institution{University of Oulu}
  \city{Oulu}
  \country{Finland}
}
\email{arif.khan@oulu.fi}

\author{Davide Taibi}
\orcid{0000-0002-3210-3990}
\affiliation{%
  \institution{University of Oulu}
  \city{Oulu}
  \country{Finland}
}
\email{davide.taibi@oulu.fi}


\begin{abstract}
Quantum computing, albeit readily available as hardware or emulated on the cloud, is still far from being available in general regarding complex programming paradigms and learning curves. This vision paper introduces \classiqnospace, a translation framework idea to bridge Classical and Quantum Computing by translating high-level programming languages, e.g., Python or C++, into a low-level language, e.g., Quantum Assembly.  Our idea paper serves as a blueprint for ongoing efforts in quantum software engineering, offering a roadmap for further \classiq development to meet the diverse needs of researchers and practitioners. \classiq is designed to empower researchers and practitioners with no prior quantum experience to harness the potential of hybrid quantum computation. We also discuss future enhancements to \classiqnospace, including support for additional quantum languages, improved optimization strategies, and integration with emerging quantum computing platforms. 
\end{abstract}

\ccsdesc[500]{Theory of computation~Design and analysis of algorithms}
\ccsdesc[500]{Theory of computation~Tree languages}

\ccsdesc[500]{Hardware~Quantum technologies}
\ccsdesc[500]{Hardware~Quantum computation}
\ccsdesc[500]{Hardware~Emerging technologies}

\keywords{Quantum Computing, Programming Languages, Quantum Programming Language, Translation, Python, QASM}

\newcommand{\classiq}{$Classi|Q\rangle$ }
\newcommand{\classiqnospace}{$Classi|Q\rangle$}


\maketitle

\section{Introduction}
\label{sec:intro}
\input{Sections/Introduction}

\section{Background \& Motivation}

\label{sec:motivation}
\input{Sections/Motivation}
\section{Design}

\label{sec:design}
\input{Sections/Design}
\section{Roadmap}
\label{sec:roadmap}
\input{Sections/Roadmap}

\section{Limitations \& Future Directions}
\label{sec:limitations}
\input{Sections/Limitations}

\section{Conlusions}
\label{sec:conlcusions}
\input{Sections/Conclusions}
 \bibliographystyle{ACM-Reference-Format} 
\setcitestyle{maxcitenames=2, mincitenames=1}
 
\bibliography{bibliography}
\end{document}

%% file: Sections/Introduction.tex
In 1938, Alan Turing introduced his theory on computable numbers and the design of the universal machine to the world, paving the way for the computing age \cite{turing1936computable}. Computing has rapidly and profoundly evolved and each passing decades marked a significant milestone in our technological journey. The advent of electronic computers led to the making of the Electronic Numerical Integrator and Computer, i.e., ENIAC, in the 1940s, which represented a monumental leap forward in processing power and computation speed \cite{hartree1946eniac}. 
The subsequent development of transistors and integrated circuits in the 1950s and 1960s paved the way for the miniaturisation of computers, making them more accessible and practical \cite{4785580}. The birth of personal computing in the 1970s and the introduction of the first commercially successful microprocessor, the Intel 4004, revolutionised how individuals interacted with technology \cite{4776530}. 
The internet era ushered in during the late 20th century, further transforming the landscape by connecting computers globally, enabling unprecedented communication and information exchange \cite{berners1992world}. Now, on the cusp of a new frontier, quantum computing (\textbf{QC}) emerges as a paradigm shift, promising to redefine the limits of computation and unravel new possibilities in the quest for computational supremacy \cite{nielsen2010quantum, DESTEFANO2022111326,10.1145/3587062.3587071}. 

Since Quantum Computing has reached the so-called quantum supremacy \cite{DBLP:journals/tse/GuoY23}, the race to “\textit{quantumize the algorithm}” is hastened by industry and research in every field \cite{KWAK2023486,DBLP:journals/qe/ZhangN20,AJAGEKAR201976,refId0,AJAGEKAR2020106630,yarkoni2022quantum,yetics2020optimization,nenno2023dynamic}.

As QC is widespread and becoming more accessible \cite{TavassoliSabzevari2024}, researchers and practitioners (\textbf{R\&Ps}) can leverage simulated and real quantum hardware for complex computational tasks \cite{doi:10.1098/rspa.2017.0551,DBLP:journals/qmi/ZhaoPRW19,DBLP:journals/corr/abs-2303-09491}.

In the 2023 IEEE Spectrum Top Programming Languages report~\footnote{\url{https://spectrum.ieee.org/the-top-programming-languages-2023}}, Python emerged as the foremost choice among researchers and practitioners, closely trailed by Java.  Python widespread adoption is attributable to its simplicity, readability, and rich library ecosystem catering to general-purpose programming and scientific computation.  Conversely, QC is still in its early stages, necessitating specialised languages like QASM or Q\#.

QASM is a low-level programming language used for specifying quantum circuits, operations, and algorithms. Similar to classical assembly languages, QASM provides a human-readable representation of quantum instructions that can be directly executed on quantum computers or simulated on classical computers. QASM allows programmers to define quantum gates, quantum measurements, and control flow operations, enabling the construction of complex quantum algorithms and experiments. It serves as an intermediary between higher-level quantum programming languages and the hardware or simulators that execute quantum computations.
Q\# is a specialized programming language tailored for quantum algorithms within the Microsoft Quantum Development Kit. Integrated with libraries and simulators, it facilitates the development of quantum applications. By leveraging a syntax akin to C\#, Q\# enables developers to simulate and execute quantum programs on classical computers. However, it lacks the capability to directly translate classical C\#-like code into pure quantum code.
In this context, allowing R\&Ps proficient with high-level languages like Python or C++ (in short, \textbf{PyC}) to harness QC power seamlessly is essential.  

Our contributions are as follows:
(\textbf{1}) We introduce \classiqnospace, our translation framework idea designed to bridge the gap between the classical and QC realms. \classiq leverages Abstract Syntax Trees (\textbf{ASTs}) as an intermediate language for source-to-source translation. 
(\textbf{2}) We introduced the newly concept of Quantum Programming Language Patterns (\textbf{QPLPs}).
(\textbf{3}) Therefore, we plan to enable \classiq to provide block-level translation leveraging newly  to replace entire blocks of PyC source code with optimised quantum code leveraging our QPLPs.  \classiq aims at seamlessly translating  PyC-based algorithms to Quantum Assembly Language (\textbf{QASM}), more specifically OpenQASM 3.0 \cite{Cross_2022}. Therefore, our contribution is twofold: we provide the first PyC to QASM translator,\classiqnospace, design and introduce the QPLP concept.

The remainder of the paper is structured as follows. We present our idea's background and motivations in Section \ref{sec:motivation}. We show the design principles in Section \ref{sec:design}. We discuss our Roadmap in Section \ref{sec:roadmap}, and address limitations in Section \ref{sec:limitations}. Finally, we draw conclusions and future research direction in Section \ref{sec:conlcusions}.

%% file: Sections/Motivation.tex
\textbf{QC current challenges.} Classical Computing (\textbf{CC}) represents information with bits that can take 0 or 1 as a state; on the contrary, QCs employ quantum bits, or qubits that leverage the principles of quantum superposition and entanglement, allowing them to exist in a combination of both states simultaneously, represented by a quantum state vector. Quantum properties enable qubits to support parallel computations; the qubit will work out all possibilities simultaneously. This Makes a huge difference in computational power. Qubits are extremely sensitive to their environment and cause decoherence, which is one of the major problems in retaining the integrity of quantum information over time. Special quantum gates are also required, quite different from the classical logic gates used with bits. Overall, qubits extend the computation capabilities from the classical bits to offer exponentially faster processing and the solution of problems intractable for classical computers, thus empowering quantum computers to accomplish specific tasks beyond the capabilities of a classical computer.\cite{9340056}. \citet{DESTEFANO2022111326}, investigated the ad of quantum programming, recognizing its evolution from a scientific interest to an industrially available technology that challenges the limits of CC. The findings revealed limited current adoption of quantum programming.

\textbf{\classiq enabling factors.} Researchers in non-computer science fields, such as physicists, mathematicians, data scientists, biologists, and many others, cannot write cumbersome and complex code such as QASM for leveraging QC powers.

ASTs and OpenQASM 3.0 are the key technologies that make \classiq development possible. ASTs are a hierarchical tree structure representing the syntactic structure of source code in a programming language. It is an abstraction of the code's syntax that disregards specific details such as formatting and focuses on the essential structure.
ASTs are a valuable tool in compiler design and programming language processing. They capture the relationships between code elements, such as expressions, statements, and declarations. By abstracting the essential components of a programming language, an AST enables tools and compilers to analyse and manipulate code in a language-agnostic manner. Therefore, \textbf{our framework can potentially convert every language to QASM}.

Similarly, OpenQASM is a specialized language crafted for quantum computer programming. It acts as a user-friendly interface, enabling the expression of quantum circuits and algorithms in a format executable by quantum processors. OpenQASM offers a structured and easily understandable method for specifying quantum gates, operations, and measurements, facilitating the creation of quantum programs for diverse quantum devices. The recent release of OpenQASM's third major version marked a significant milestone by introducing classical control flow, instructions, and data types. As the OpenQASM specification outlines, this addition allows for defining circuits involving real-time computations on classical data, \textbf{paving the way for genuinely hybrid solutions and serving as a driving force of \classiqnospace, motivating and empowering it}. Therefore, \classiq serves as the pivotal tool for multidisciplinary collaboration. Our framework seamlessly translates PyC-based algorithms into QASM, enabling quantum-hybrid computation. The translator accelerates research and innovation by overcoming language barriers and CC knowledge, allowing teams to explore novel applications while accommodating subteams' preferred languages.




Different backgrounds and terminologies used in quantum technologies are a major problem highlighted by, e.g. \citet{10.1145/3587062.3587071}. The complexity of quantum algorithms makes them inaccessible for most of current classical developers. In this respect, \citet{10.1145/3587062.3587071} have pointed out that bridging knowledge gaps among disciplines will address these problems. Therefore, technical expertise, interdisciplinary collaboration, as well as knowledge integration, play crucial roles in both QC as well as software engineering. Our work stems from these challenges, presenting a possible bridge for R\&P to harness QC power more efficiently. The programming languages and resource power limitations of state-of-the-art QC providers have led to a preference for hybrid approaches surpassing classical computational power \cite{AJAGEKAR2020106630}. Despite prevailing research trends, \classiq seeks to bridge the gap between proficient classical practitioners and the QC potentials, presenting a seamless transition path. 

\classiq is tailored to a diverse team; to our knowledge, such a framework was never developed. Leveraging Python's simplicity, the code is easily understood by researchers with low coding proficiency yet robust enough to optimize recurrent computational patterns when translating to QASM.
\begin{figure*}
\centering
\begin{minipage}[b]{.4\textwidth}

    \includegraphics[width=0.85\linewidth]{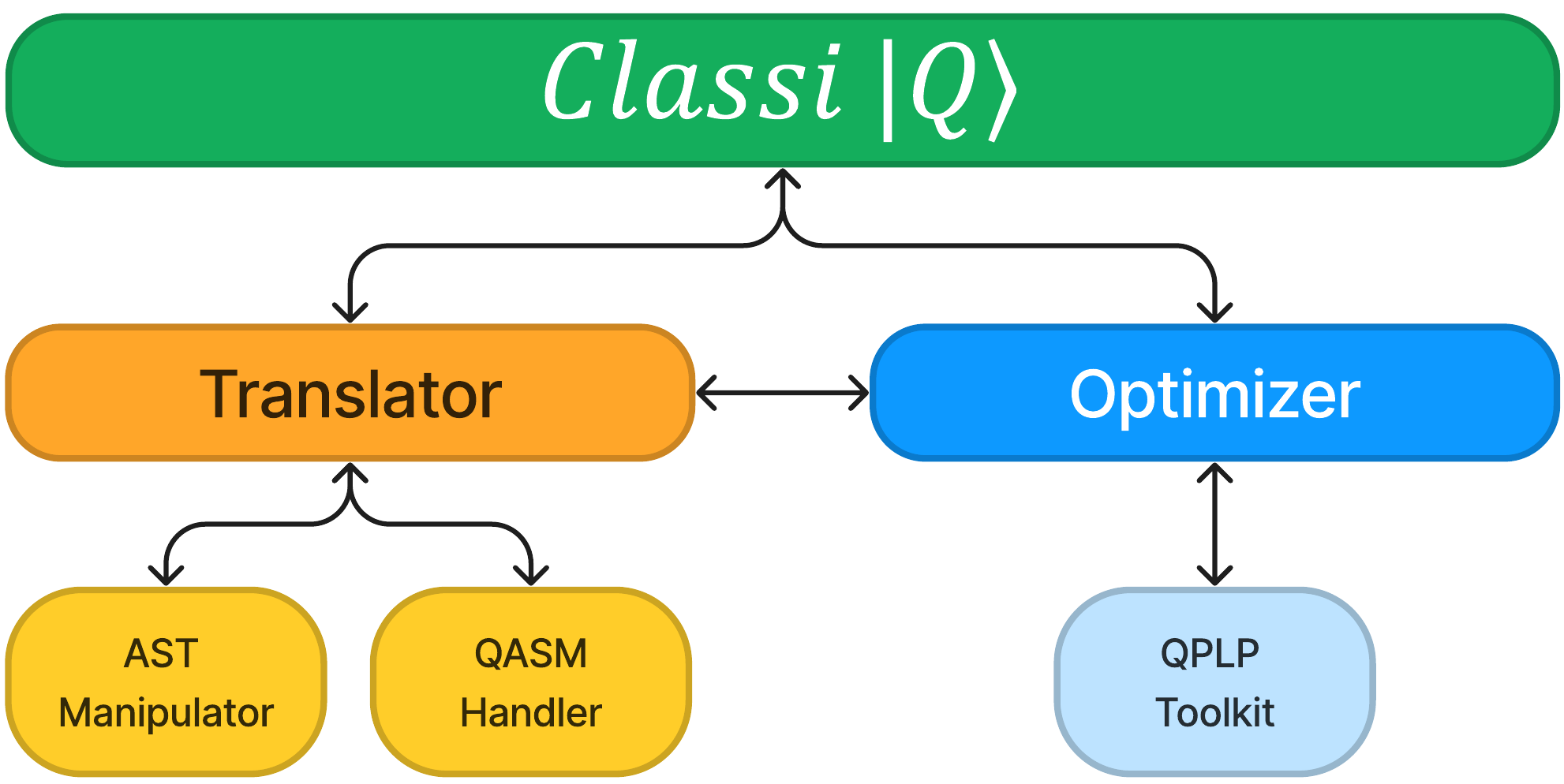}
    \caption{Overview of the \textbf{\classiq} Framework}
    \label{fig:framework}

\end{minipage}\qquad
\begin{minipage}[b]{.4\textwidth}
    \includegraphics[width=0.85\linewidth]{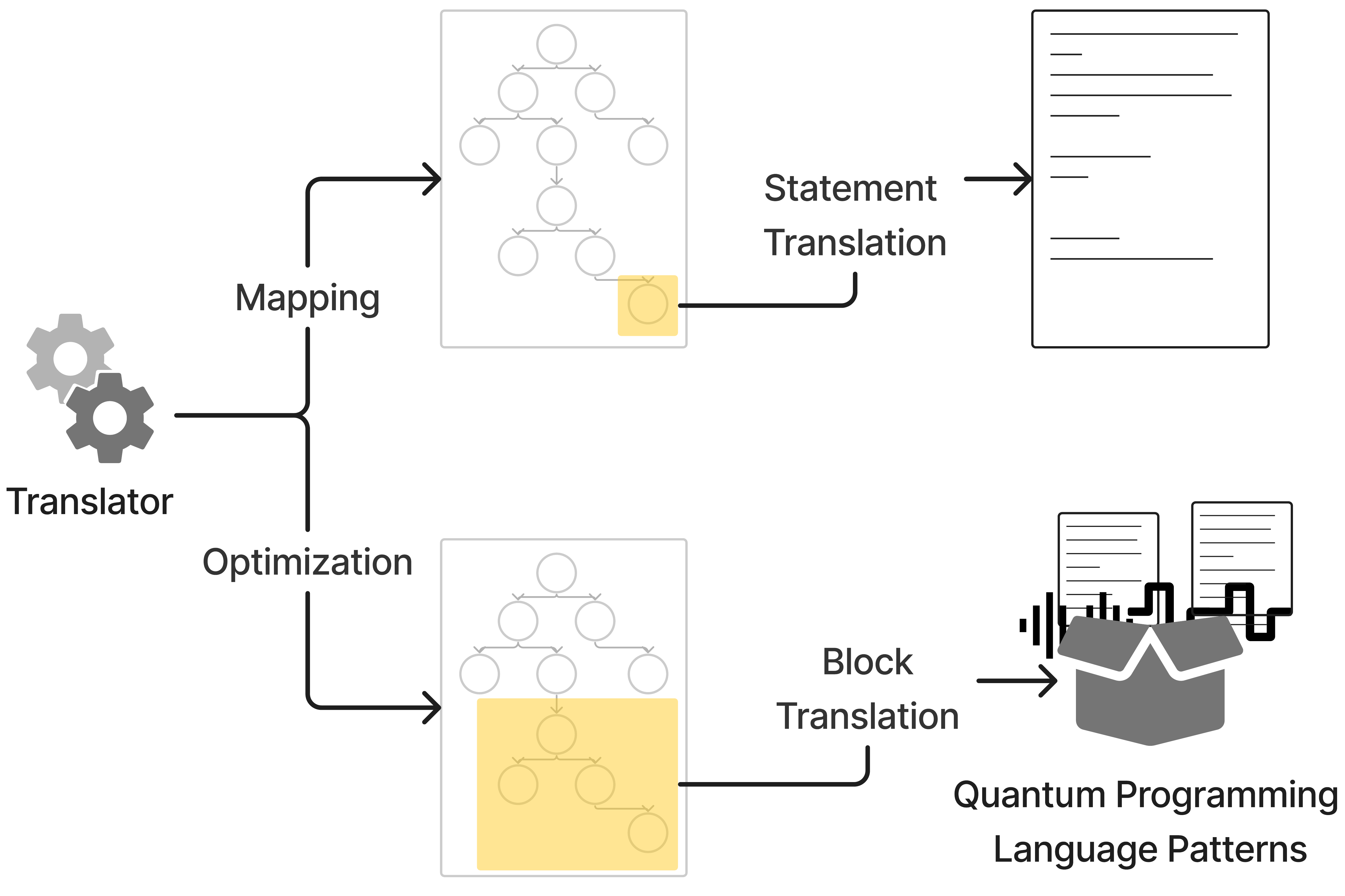}
    \caption{Workflow Overview}
    \label{fig:workflowOverview2}
\end{minipage}
\end{figure*}

%% file: Sections/Design.tex
In this section, we introduce the concept of QPLPs and the design of \classiq.
\subsection{Quantum Programming Language Patterns }

Recent studies, introduced patterns in QC realm \cite{DBLP:conf/kivs/Leymann17,Buehler2023_QuantumSoftwareEngineeringPatterns, Georg2023_PatternsQuantumExecution}. For instance, \citet{DBLP:conf/kivs/Leymann17} propose patterns for handling essential operations within the QC realm, such as state preparation, entanglement and unentanglement, phase shift, and many others. In the same vein, \citet{Georg2023_PatternsQuantumExecution} presents patterns focused on the execution stage while \citet{Buehler2023_QuantumSoftwareEngineeringPatterns} showcased development-pattern such as: Quantum Module, Quantum Module Template, and Quantum Circuit Translator, a pattern for translating a circuit among different QC vendors. R\&P can find the previous patterns in the quantum computing patterns online library\footnote{\url{https://quantumcomputingpatterns.org/}}. 

\citet{gamma1995design} pioneered the concept of patterns in software engineering. The motivation behind their book was to leverage the power of object-oriented programming and provide reusable elements that would make software more flexible, modular, and reusable. We want to define the QPLPs as \textit{elements of reusable QC algorithm}. We envision collecting or developing patterns that allow R\&P to replace entire blocks of classical computation, such as the computation of the discrete algorithm \cite{DBLP:journals/jmc/Ekera21} or a mean \cite{DBLP:journals/corr/abs-1909-11485}, the search for an element in an array \cite{DBLP:conf/stoc/Grover96}, with black-quantum boxes that leverage the full power of the underlying quantum hardware. The optimization module will then leverage our collection of QPLPs and replace CC blocks with QC-powered blocks.
To our knowledge, none of the existing patterns consider programming language patterns. We can trace this research gap to two root causes: 1) each QC vendor proposes its implementation and design of a language that leads to 2) scarce interoperability among vendors. In this scenario, the most promising QC language is QASM \cite{DBLP:conf/kivs/Leymann17, Cross_2022}, and it can have the impact that Java had when first released.

\subsection{\classiq}

Figure \ref{fig:framework} shows the building block of \classiq. The framework comprises two main blocks: the \textbf{Translator} and the \textbf{Optimizer}. 
The translator module (\textbf{TM}) handles the translation of CP via its AST representation. Within TM, the AST manipulator sub-module handles AST operations like tree traversal and source code statement interpretation. The QASM Handler sub-module leverages a custom grammar, i.e., mapping file, to interpret and translate the abstract representation of the code into the corresponding OpenQASM equivalent. 

The optimizer module (\textbf{OM}) optimizes entire code blocks. We design OM to analyze multiple statements, i.e., blocks, of code and leverage QPLPs to replace the entire block with a predefined quantum subroutine.

More specifically, Figure \ref{fig:workflowOverview2} presents the overview of the translation workflow. \classiq will enable R\&P to translate CP with our without optimizations. In the "mapping translation" scenario, \classiq transverses the AST, analyzes its content, and translates each statement in the corresponding OpenQASM classical representation. On the other hand, block translation aims to translate entire source code blocks, replacing their content with an improved quantum algorithm. This block replacement stems from our definition of QPLPs. Hence, it will read the source code, understand what kind of computation the R\&P envisioned or requested via custom notation, and provide one or more possible alternatives to compute the same output with the improved QC algorithm version.

%% file: Sections/Roadmap.tex
This section briefly presents the roadmap we will pursue to bring \classiq to fruition. We are developing the TM and the OM functionalities in parallel. Specifically, we target Python as the first proof of concept (\textbf{POC}). We believe to be able to release the first beta of the TM by the end of summer 2024.
\textbf{The translation module} is the core of \classiq basic functionalities; hence, it is our main focus. We finished the development of the AST transversal for Python and are currently implementing the statement translation leveraging the grammar provided by OpenQASM.
We are currently exploring the literature and experimental implementations in detail to identify QPLPs suitable for classical sub-problems within the quantum algorithm. Once promising patterns are identified, the focus will shifts to seamlessly incorporating them into the existing translation workflow in the \textbf{QPLP toolkit} sub-module.

%% file: Sections/Limitations.tex
This section acknowledges limitations to our idea and possible future directions.

\classiq is designed to support PyC grammars and syntaxes comprehensively. However, it is essential to note that object translation remains unavailable at present. This limitation stems from the intrinsic nature of QASM as a "low-level" language, lacking support for high-level constructs such as objects. Nevertheless, we will support all built-in functions, enabling the translation of complex calculations by composing programs exclusively with built-in functions. Future works should address this limitation through smart recursive translation or expanding the QASM capabilities, which is essential for enhancing the hybrid approach.

While the current focus of translation is on Python and C++, it is noteworthy that \classiq will ultimately evolve into a source-language-agnostic tool. Consequently, with customized grammar, it can translate other source languages, such as Q\# or Java, to QASM.

Finally, since our vision emphasizes translating code from classical programming languages, future research should focus on investigating potential issues that may inadvertently arise in the quantum domain from this translation, such as defects and vulnerabilities \cite{esposito2023uncovering, esposito2024validate}. For example, there are currently no definitions of quantum vulnerabilities in terms of Common Weakness Enumeration (CWE) \cite{esposito2023can}, nor is there an accepted definition of quantum vulnerability severity. Therefore, future research efforts should focus on developing static analyzers to detect these new types of defects and vulnerabilities, and to assess their severity\cite{esposito2024extensive}.

%% file: Sections/Conclusions.tex
In this work, we have introduced \classiqnospace, the design of the ideal PyC to QASM translator, bridging the classical and quantum realms. To illustrate the practical relevance of \classiqnospace, we outline three hypothetical scenarios grounded in real-world contexts. Moreover, we detail the design, key decisions, and the introduction of QPLPs. The ultimate goal of \classiq is to empower researchers and practitioners to exploit the capabilities of quantum computing without requiring specific training.